\documentclass[AMA,STIX1COL]{WileyNJD-v2}
\usepackage{graphicx}
\usepackage{placeins}
\usepackage{multirow}

\newcommand{\df}{\mbox{\textit{df}}}

\newcommand{\rb}[1]{\raisebox{1.5ex}[-1.5ex]{#1}}

\def\btheta{{\boldsymbol\theta}}
\def\bepsilon{{\boldsymbol\epsilon}}
\def\bSigma{{\boldsymbol\Sigma}}
\def\bDelta{{\boldsymbol\Delta}}

\def\bmu{{\boldsymbol\mu}}
\def\bX{{\mathbf X}}

\def\bW{{\mathbf W}}

\def\bC{{\mathbf C}}
\def\bB{{\mathbf B}}
\newcommand{\bfdelta}{\mbox{\boldmath $\delta$}}
\newcommand{\bbeta}{\mbox{\boldmath $\beta$}}
\newcommand{\MSE}{\mbox{MSE}}
\newcommand{\CP}{\mbox{CP}}

\articletype{Article Type}%

\received{xx}
\revised{xx}
\accepted{xx}

\begin{document}

\title{Model selection for component network meta-analysis in
  connected and disconnected networks: a simulation study}

\author[1]{Maria Petropoulou}

\author[1]{Gerta Rücker}

\author[2]{Stephanie Weibel}

\author[2]{Peter Kranke}

\author[1]{Guido Schwarzer}

\authormark{PETROPOULOU \textsc{et al}}

\address[1]{\orgdiv{Institute of Medical Biometry and Statistics},
  \orgname{Faculty of Medicine and Medical Center – University of
    Freiburg}, \orgaddress{\state{Freiburg}, \country{Germany}}}

\address[2]{\orgdiv{Department of Anaesthesiology, Intensive Care,
    Emergency and Pain Medicine}, \orgname{University of Würzburg},
  \orgaddress{\state{Würzburg}, \country{Germany}}}

\corres{Maria Petropoulou, Institute of Medical Biometry and
  Statistics, Faculty of Medicine and Medical Center – University of
  Freiburg, Zinkmattenstraße 6A, 79108 Freiburg,
  Germany. \email{petropoulou@imbi.uni-freiburg.de}}


\abstract[Abstract]{
Network meta-analysis (NMA) is widely used in evidence synthesis to estimate the effects of several competing interventions for a given clinical condition. One of the challenges is that it is not possible in disconnected networks.

Component network meta-analysis (CNMA) allows technically “reconnecting” a disconnected network with multicomponent interventions. The additive CNMA model assumes that the effect of any multicomponent intervention is the additive sum of its components. This assumption can be relaxed by adding interaction component terms, which improves the goodness of fit but decreases the network connectivity. Model selection aims at finding the model with a reasonable balance between the goodness of fit and connectivity (selected CNMA model).

We aim to introduce a forward model selection strategy for CNMA models and to investigate the performance of CNMA models for connected and disconnected networks. We applied the methods to a real Cochrane review dataset and simulated data with additive, mildly, or strongly violated intervention effects. We started with connected networks, and we artificially constructed disconnected networks. We compared the results of the additive and the selected CNMAs from each connected and disconnected network with the NMA using the mean squared error and coverage probability.

CNMA models provide good performance for connected networks and can be an alternative to standard NMA if additivity holds. On the contrary, model selection does not perform well for disconnected networks, and we recommend conducting separate analyses of subnetworks.
}

\maketitle

\section{Introduction}\label{sec1}

Standard network meta-analysis (NMA) synthesizes direct and indirect
evidence of randomized controlled trials (RCTs) to
estimate the effects of several competing
interventions.\cite{Sala:Higg:Ades:Ioan:eval:2008, Sala:indi:2012,
  Efthi:Debra:Valke:2015} One typical requirement is that the network
of interventions is connected.  However, in practice, many situations
can lead to disconnected networks with two or more subnetworks when
synthesizing evidence from RCTs.

Standard NMA is not possible in disconnected networks, instead, it can
be simply replaced with separate NMA analyses for each of the
subnetworks. Several alternative NMA methods have been proposed to
deal with disconnected networks.\cite{stevens2018} A theoretical
framework of the Bayesian contrast-based model has recently been
provided for disconnected networks.\cite{beliveau2021} Arm-based
approaches have also been developed to analyse connected and
disconnected networks.\cite{Hawki:Scott:Woods:2016,
  Hongh:Chuha:Zhang:2016, Pieph:Madde:Roger:2018} Goring et
al.\ \cite{Gorin:Gusta:Liuy:2016} proposed the random baseline
treatment effects NMA model in the Bayesian framework to accommodate
disconnected networks, while B\'eliveau et
al.\ \cite{Beliv:Gorin:Platt:2017} conducted a case study to evaluate
the performance of the random baseline treatment effects in
disconnected networks. Mixture models of RCTs and observational
studies can ``reconnect'' disconnected networks using
matching-adjusted indirect comparisons
\cite{Petto:Kadzi:Brnab:Saure:Belge:MAIC:2019, Schmitz2018a,
  Phill:Mdavi:Ades:2017, signorovitch2010, Sign:MAIC:2012,
  Veroniki2016a} or hierarchical models.\cite{Thom2015a} Linking
disconnected networks can also be addressed through
dose-response\cite{Pedder:2021} or component
NMA\cite{Rucke:Petro:Schwa:2018, Welto:Jnick:Caldw:2009} if
subnetworks share common intervention doses or components.

Many healthcare interventions consist of multiple, possibly
interacting, components. Several meta-analytical models
address the effects of such complex
interventions.\cite{Petro:Efthi:Salan:2021} Component network
meta-analysis (CNMA), a generalization of standard NMA, estimates the
effects of components of interventions and allows ``reconnecting'' a
network if there are common components in different
subnetworks.\cite{Rucke:Petro:Schwa:2018, Welto:Jnick:Caldw:2009} A
CNMA model assumes that the effect of any combination of components
is the additive sum of their components, known as the additivity
assumption. This assumption can be relaxed by adding interaction terms
allowing components of complex interventions to interact, either
synergistically or antagonistically. Adding interactions might improve the
the goodness of fit, but also decrease the network connectivity. Model
selection aims at finding a model with a reasonable balance between
the goodness of fit and connectivity (selected CNMA). The forward model
selection for CNMA models, which has recently been developed,
\cite{Rucke:Schmi:Schwa:2020} starts with a sparse (additive) CNMA
model and, by adding interaction terms, ends up with a rich CNMA.

At the moment, there is no established guidance concerning which
CNMA approach (a sparse or rich version) fits best under different
circumstances for connected or disconnected networks. Therefore, we
conducted a comprehensive simulation study to investigate
whether the CNMA model from the model selection process can be a
reasonable approach to deal with disconnected networks and an
appropriate alternative to standard NMA for connected networks. In the
simulations, we assumed that additivity either holds or is mildly or
strongly violated for one combination. Starting with a connected
network, we artificially constructed disconnected networks and
implemented the forward selection strategy to select the best CNMA
model for each disconnected network.

For connected networks, we compared mean square errors and coverage
probabilities of the selected CNMA model with those of the standard
NMA and additive CNMA model. We investigated the circumstances under
which a sparse (e.g., additive) or a richer (e.g, selection-based)
CNMA model is preferable to the standard NMA. For disconnected
networks, we compared the results of the selected CNMA model with the
additive model. We applied model selection also to data of a Cochrane
review on postoperative nausea and
vomiting.\cite{Weibel:2020a,Weibel:Schaefer:Raj:2020}

The paper is organized as follows: Section \ref{data} introduces the
motivating example; Section \ref{nmas} describes the (C)NMA models
evaluating the effects of complex interventions and the CNMA model
selection method; Section \ref{simdesign} summarizes the design of our
simulation study; Section \ref{sec:results} presents the results of the
motivating example and outlines the results of our simulation study
and, finally, Section \ref{con} discusses the main findings of the
study.

\section{Example: Cochrane review on postoperative nausea and vomiting}\label{data}

\begin{figure}[tb]
\centerline{\includegraphics[width=12cm,angle=0]{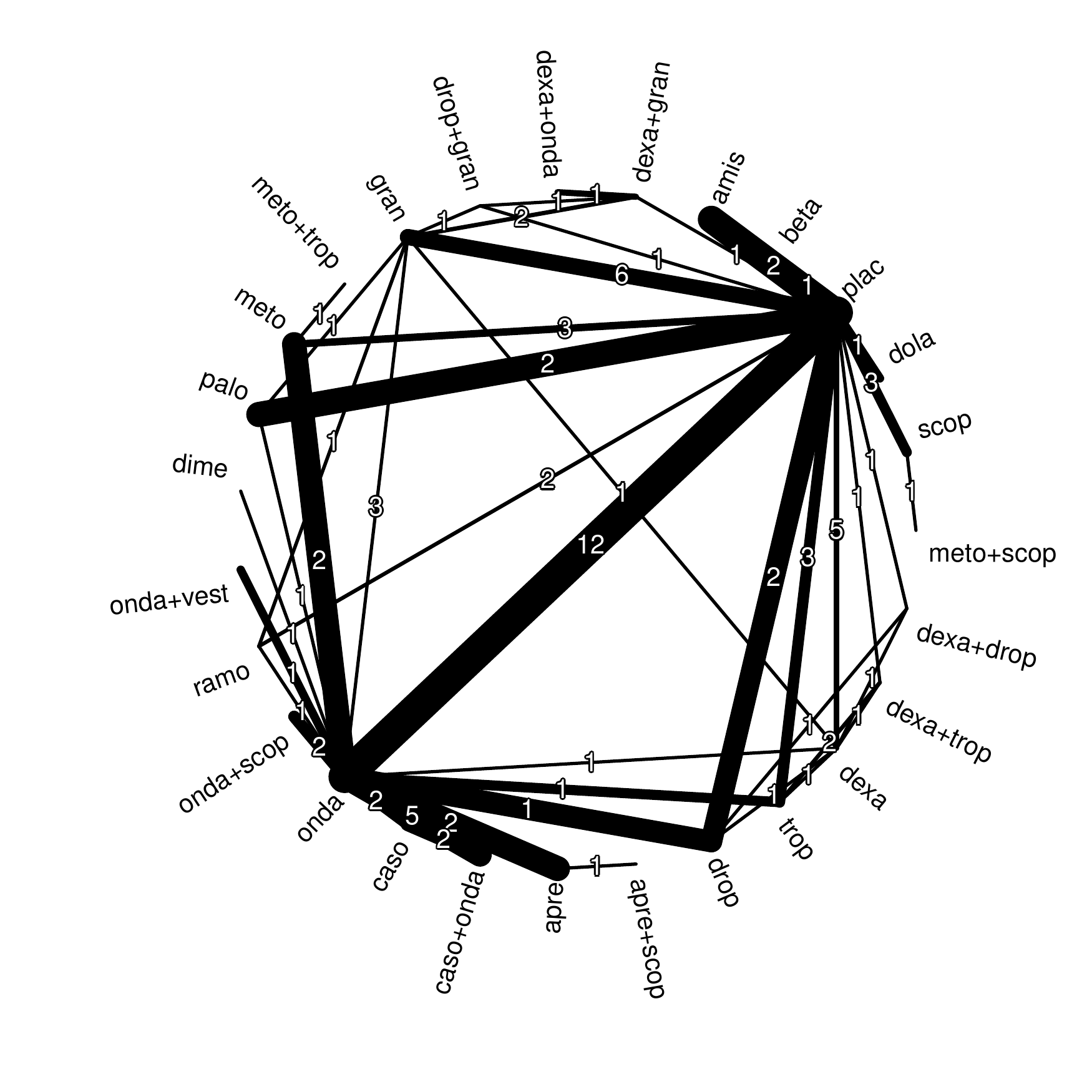}}
\caption{Network plot for the Cochrane data set (outcome: any adverse event). Line width corresponds to inverse of random effects
  standard error. Abbreviations: amis: amisulpride; apre: aprepitant;
  beta: betamethasone; caso: casopitant; dexa:dexamethasone; dime:
  dimenhydrinate; dola: dolasetron; drop: droperidol; gran:
  granisetron; meto: metoclopramide; onda: ondansetron; palo:
  palonosetron; plac: placebo; ramo: ramosetron; scop: scopolamine;
  trop:tropisetron; vest: vestipitant. Numbers represent the number of
  studies.}\label{fig:networkplot}
\end{figure}

We used a published Cochrane review of $585$ RCTs that compares
interventions for postoperative nausea and vomiting in adults after
general anaesthesia.\cite{Weibel:2020a,Weibel:Schaefer:Raj:2020} Here
we consider the outcome of any adverse event which was available in
$61$ RCTs of which four RCTs without any adverse event were excluded. The
relative effects were measured as risk ratios (RRs). In total, $27$
interventions are compared, including 15 single interventions (e.g.,
ondansetron (onda), scopolamine (scop)), 11 combinations of
interventions (e.g., ondansetron plus scopolamine (onda+scop)), and
placebo (Figure \ref{fig:networkplot}). The interventions contain 17
components (including placebo), one component (vest) was only
evaluated in a combination: onda + vest. The authors conducted
standard NMA and CNMA analyses to assess the effects of the several
competing interventions and their combinations. As Figure
\ref{fig:networkplot} shows, the network of interventions is
connected. A total of 16 interventions are compared directly with
placebo (e.g., dolasetron (dola) versus placebo).

\section{Methods}\label{nmas}

\subsection{Standard NMA}

Standard NMA assumes that each (single or combined) intervention has
its own effect which is represented as a node in the network. We
follow the frequentist approach introduced by R\"ucker
(2012).\cite{Rcke:netw:2012} Suppose we have data consisting of $m$
pairwise comparisons with $n$ interventions, and let $\btheta$
represent the $n$ intervention-based (true) responses. Let
$\mathbf{d}= (d_{1}, d_{2},..., d_{m})$ be the observed (relative)
intervention effects with the associated standard error $\mbox{SE}
(d_j)$ for each comparison $j = 1, \dots, m$.  Assuming a common
between-study variance (heterogeneity $\tau^2$) across the pairwise
comparisons, the random-effects network meta-analysis model is
\begin{equation*} 
\mathbf{d} = \bX\btheta + \bmu + \bepsilon,
\bepsilon \sim \mathcal{N}(\mathbf{0},\bSigma),
\bmu\sim \mathcal{N}(\mathbf{0},\bDelta)
\end{equation*}

where $\bX$ is the $m \times n$ design matrix describing the network
structure, $\bSigma$ is the within-study variance-covariance matrix,
and $\bDelta$ is the between-study variance-covariance matrix. Let
$\bW$ be a diagonal $m \times m$ weight matrix with a vector of
weights on its diagonal. The weight for each two-arm study is the
inverse of the sum of the within- and between-study variance. For
multi-arm studies, the weights are assumed to be adjusted as described
in R\"ucker and Schwarzer (2014).\cite{Rucke:Schwa:2014} We can write
the standard NMA model briefly $\bfdelta = \bX\btheta$ where
$\bfdelta$ denotes the vector of true relative intervention effects
which is estimated using weighted least squares regression
($\hat\bfdelta$). Cochran's $Q$ statistic is given by $Q = (\mathbf{d}
- \hat\bfdelta) ^\top \bW (\mathbf{d} - \hat \bfdelta)$ which follows
a chi-square distribution with degrees of freedom $\df
= n_a - k - (n - 1)$, where $n_a$ is the total number of intervention
arms and $k$ is the number of studies. More details for the model can
be found in R\"ucker et al.\ \cite{Rcke:netw:2012, Rucke:Schwa:2014}

\subsection{Additive CNMA}

The sparse additive CNMA model assumes that the effect of each
combined intervention is the additive sum of the effects of its components, that is,
equal components cancel out in pairwise
comparisons.\cite{Rucke:Petro:Schwa:2018, Welto:Jnick:Caldw:2009} Let
the number of components be $c$. Having the data consisting of $m$
pairwise comparisons with $n$ interventions, the design matrix of the
additive CNMA model is the $m \times c$ matrix given by $\bX_{a}
=\bB\bC$ , where $\bB$ is the $m \times n$ design matrix describing
which interventions are compared in each pairwise comparison, and
$\bC$ is the $n \times c$ combination matrix describing the
information on how the $n$ interventions are composed of the $c$
components.\cite{Rucke:Petro:Schwa:2018} The additive CNMA model is
given by
\begin{equation} \label{additive} 
  \bfdelta_{a} = \bX_{a} \bbeta = \bB \bC \bbeta = \bB \btheta_{a}
\end{equation}
where $\bfdelta_{a} \in \mathbb R^m$ is the vector of true relative
intervention effects, $\bbeta \in \mathbb R^c$ a parameter vector of
length $c$, representing the component effects, and $\btheta_{a} = \bC
\bbeta \in \mathbb R^n$ a vector of length $n$, representing the
intervention effects.

\subsection{Interaction CNMA}

The interaction CNMA model is an extension of the additive CNMA
model.\cite{Rucke:Petro:Schwa:2018, Welto:Jnick:Caldw:2009} The model
assumes an interaction between two or more observed components
(antagonistically or synergistically) and therefore the combination of
components provides larger or smaller effects than the sum of their
effects, respectively. The interactions of interest can be added as
additional columns to the combination matrix
$\bC$.\cite{Rucke:Petro:Schwa:2018} For $l$ interactions, the
combination matrix $\bC_{int}$ is of dimension $n \times (c+l)$. An
interaction CNMA model is implemented in complete analogy to the
additive CNMA model. Therefore, having the design matrix $\bX_{int}
=\bB\bC_{int}$, the interaction CNMA model is given by
\begin{equation} 
  \bfdelta_{int} = \bX_{int} \bbeta_{int} = \bB \bC_{int} \bbeta_{int}
  = \bB \btheta_{int}
\end{equation}
where $\bfdelta_{int} \in \mathbb R^m$ is the vector of true relative
intervention effects, $\bbeta_{int}$ a parameter vector of length
$c+l$, representing the component and interaction effects, and
$\btheta_{int} = \bC_{int}\bbeta_{int} \in \mathbb R^n$ a vector of
length $n$, representing the intervention effects.

$\hat\bfdelta_{a}$ from the additive and $\hat\bfdelta_{int}$ from the
interaction model are estimated using weighted least squares
regression. Details on the estimation and the multivariate version of
Cochran's Q for CNMA models can be found in R\"ucker et
al.\ \cite{Rucke:Petro:Schwa:2018}

\subsection{CNMA model selection \label{sec:cnma:selection}}

Table \ref{tab:fit} illustrates the balance between the goodness of fit
and connectivity in (C)NMA models for (dis-)connected networks. For
connected networks, the standard NMA model is the richest model with
the smallest $\df$, its $Q$ is the smallest under all models, and it
thus fits the data better than other models, though not necessarily
well as heterogeneity and inconsistency may be present. On the other
hand, the additive model is the most parsimonious (sparse) model,
i.e., it has the smallest number of parameters, the largest $\df$, and
the largest $Q$. On the downside, model fit, measured by Cochran’s
$Q$, is poor if the additivity assumption is violated. Adding
interactions typically decreases $Q$, thus improving the goodness of
fit, but also decreases the degrees of freedom and, most notably, may
decrease the network connectivity ($\df\searrow$)(Table
\ref{tab:fit}). In the supplement, we describe an algorithm to identify
inestimable interactions.

R\"ucker et al.\ \cite{Rucke:Schmi:Schwa:2020} introduced a model
selection procedure for CNMA models with the aim to find the CNMA
model with a reasonable balance between the goodness of fit and
connectivity. There are two possible directions: forward selection and
backward selection.

Forward CNMA model selection starts with the additive (sparse) CNMA
model and is moving forward to richer models. During the selection
process, observed interactions are gradually added to the model until
a stopping criterion is fulfilled.\cite{Rucke:Schmi:Schwa:2020} We use
the Akaike Information criterion (AIC) as a stopping
criterion.\cite{heinze2018} The selection process stops if all
p-values of the remaining estimable interactions are above 0.157.

Backward CNMA model selection should start with an interaction CNMA
model having the same value for Cochran's $Q$ as the standard (rich)
NMA model. During the selection process, one interaction term is
removed from the model in each step until a stopping criterion is
fulfilled.\cite{Rucke:Schmi:Schwa:2020} One difficulty in connected
networks is to determine a sufficient number of interactions to get
the same value for Cochran's $Q$. In disconnected networks, a standard
NMA is impossible and in this case, additivity can be assumed for just
one component that is common to all
subnetworks.\cite{Rucke:Schmi:Schwa:2020} Therefore, we may start with
‘separating’ one component which is common to all subnetworks.
\cite{Rucke:Schmi:Schwa:2020} Such a model usually provides a good
fit, it may even provide the minimum $Q$, given by the sum over all
$Qs$ from the subnetworks. However, it provides only a very loose
connection between the subnetworks and is associated with small
$\df$ (Table \ref{tab:fit}). For this reason, under all
models giving the same model fit, we prefer those with greater
connectivity than those with smaller connectivity. In other words, we
prefer sparse models to rich models.

\begin{table}[htb!]
  \caption{(C)NMA models for (dis-)connected networks. $n_{c} = $ number
  of subnetworks (connectivity components), $n_{a} =$ number of
  intervention arms, $k =$ number of studies, $n= $ number of
  interventions, $r=$ rank of the design matrx of additive CNMA model.
} \label{tab:fit}
\vspace*{0.5cm} \centering
\begin{tabular}{ccccc} 
  & & & Connected network & Disconnected network \\
  \hline 
   & \multicolumn{2}{c}{CNMA models}          & NMA  & Separate NMAs       \\
   & Additive model &  Interaction models     &      & for each subnetwork \\ \hline 
  No. of interactions & none & 1, 2, 3, \dots & all observed & all observed \\ \hline
  Model fit      & often poor fit &  $\nearrow$ & often good fit & maximal fit \\
  $Q$            & &  $\searrow$ & & minimal $Q=\sum_{i = 1}^{n_c}Q_i$\\
  \hline
  Connectivity & good connectivity&  $\searrow$&  poor connectivity & minimal connectivity\\
  $\df$  & maximal $\df=n_a - k - r $ & $\searrow$& $\df=n_a - k - (n - 1)$ & minimal $\df=n_a - k - (n - n_c)$\\
  \hline
\end{tabular}
\end{table}

\subsection{Construction of disconnected networks \label{sec:disc:generate}}

As described in the previous section, CNMA model selection can be
applied to connected and disconnected networks. To evaluate
model selection in disconnected networks, we artificially constructed
disconnected networks in the Cochrane data set and simulation study
following an approach similar to B\'eliveau et
al. \cite{Beliv:Gorin:Platt:2017}

Having a connected network with $n$ interventions, we constructed
disconnected networks with two or three separate subnetworks without
dropping any interventions. We started by constructing a \textit{main
  subnetwork} which includes the reference intervention, i.e., placebo
in the Cochrane data set. All interventions that are connected to the
reference only by a direct comparison have to be part of the main
subnetwork. If such a comparison was part of a multi-arm study, then
all interventions evaluated in this multi-arm study were also part of
the main subnetwork. For example, amis, beta and, dola must be part of
the main subnetwork in the Cochrane data set as they are only compared
with placebo. Otherwise, these interventions could not be included in
the disconnected network and the network size would be smaller than
$n$. We call the interventions which must be part of the main
subnetwork the \textit{minimal set} of interventions. In the Cochrane
data set, the minimal set consists of 17 interventions if the placebo is
used as reference (Table A1).  Interventions that are connected to the
reference via a loop may be added to the minimal set to construct a
disconnected network as long as a second subnetwork exists. In fact,
well-connected networks may not contain a single intervention that
must be part of the minimal set. In this case, the main subnetwork is
constructed by adding at least one intervention to the minimal set
consisting of the reference only.
 
All interventions not included in the main subnetwork are part of the
\textit{auxiliary subnetwork(s)}. A disconnected network is
constructed by removing all studies comparing interventions from
different subnetworks. In the Cochrane data set, 19 studies with 33
pairwise comparisons have to be removed from the data set in order to
separate the minimal set (with 17 interventions, 24 studies, 38
pairwise comparisons) from the auxiliary network (10 interventions, 14
studies, 18 pairwise comparisons). Other disconnected networks can be
constructed by adding interventions and comparisons to the minimal
set. We wrote an R script to identify all disconnected networks for a
given minimal set (see GitLab repository
\url{https://gitlab.imbi.uni-freiburg.de/sc/cnma-simulation}).

\section{Simulation design}\label{simdesign}

\subsection{Simulation scenarios}

\begin{figure}[htb!]
\centerline{\includegraphics[width=12cm,angle=0]{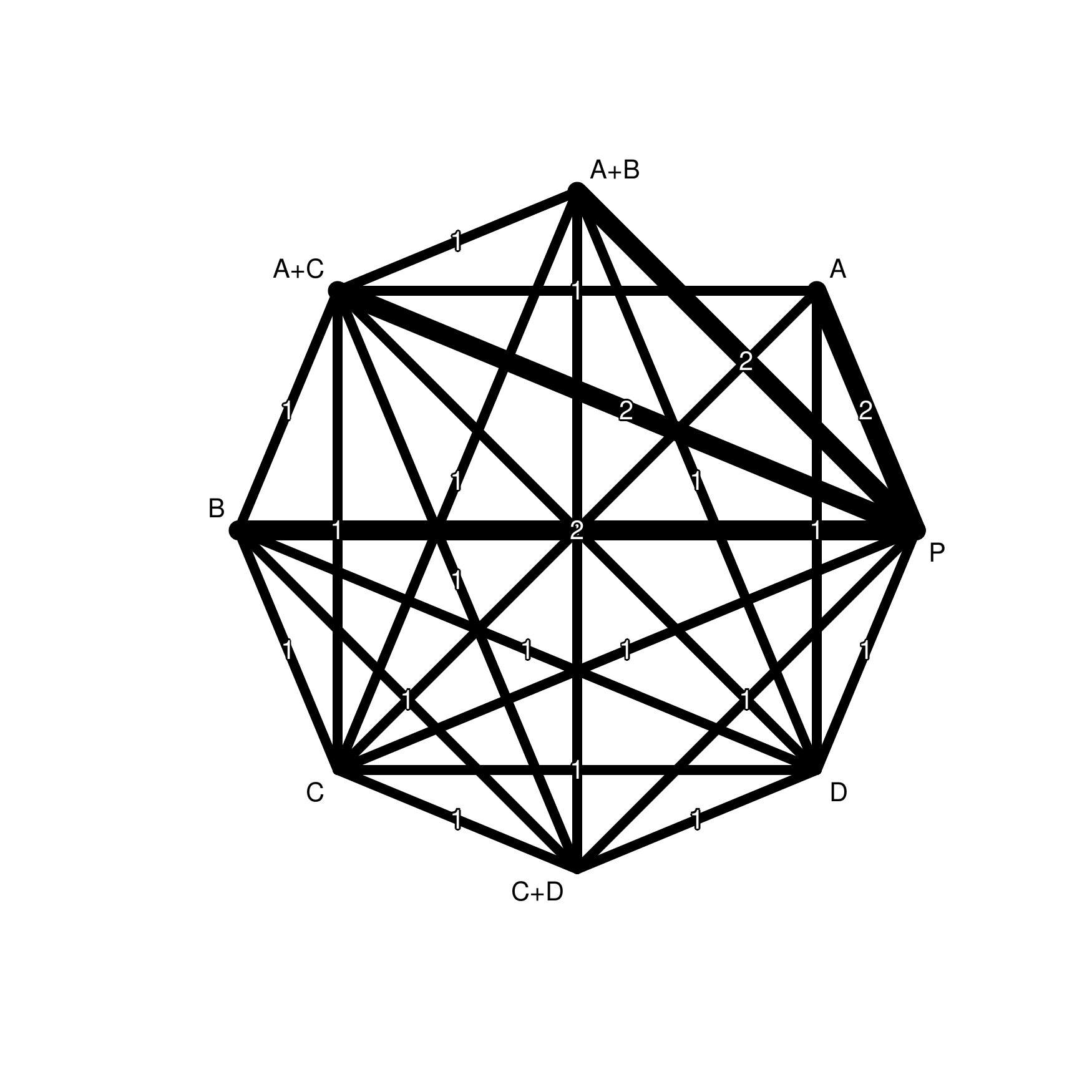}}
\caption{Network plot of simulated network.  Line width corresponds to
  numbers of studies included in the
  comparison.\label{fig:networkplot-art}}
\end{figure}

We simulated data for a network of two-arm studies with eight
interventions ($n = 8$): four single treatments $(A, B, C, D)$, three
combinations $(A+B, A+C, C+D)$ and placebo $P$. The network is
well-connected, however, omits the direct comparisons $A$ versus $B$,
$A$ versus $A+B$ and $A$ versus $C+D$ (Figure
\ref{fig:networkplot-art}). We assumed two studies directly comparing each of
interventions $A, B, A+B, A+C$ with placebo (which was chosen as the
reference) and only a single study for other
comparisons. We generated arm-level dichotomous outcome data with odds
ratio as effect measure and we assumed a common heterogeneity
variance $\tau^2$ for all pairwise comparisons. Non-equal true
relative effects ($\bfdelta=\log(\mbox{OR})$) were set with
$e^{\delta_{A,P}} = 1.40$, $e^{\delta_{B,P}} = 1.20$,
$e^{\delta_{C,P}} = 2.30$ and $e^{\delta_{D,P}} = 1.50$. A summary of
all simulation parameters is given in Table \ref{tab:simuset}.

Starting with the connected network in Figure
\ref{fig:networkplot-art}, we artificially constructed disconnected
networks differing in network geometry, number of included studies,
and pairwise comparisons (see Section \ref{sec:disc:generate}). The
forward model selection strategy described in Section
\ref{sec:cnma:selection} was implemented and the selection-based CNMA
model was chosen for each disconnected network.

Following Thorlund and Mills \cite{Thorl:Mills:2012}, we were
interested in three scenarios for intervention effects:

\begin{itemize}
\item[(a)] \textit{All relative intervention effects are
  additive} \newline For any two interventions, say $A$ and $B$, the relative
  effect of the intervention $A+B$ comprising two components $A$ and
  $B$ versus intervention $P$ is the additive sum of the relative
  intervention effect of $A$ versus $P$ and the relative intervention
  effect of $B$ versus $P$:
\begin{equation*} 
  \delta_{A+B,P} = \delta_{A,P} + \delta_{B,P}
\end{equation*}

\item[(b)] \textit{The additivity assumption for one intervention is
  mildly violated with a relevant synergistic interaction} \newline
  Assuming an interaction ratio $\mbox{\textit{IR}}_{AB}$ between
  interventions $A$ and $B$, the relative intervention effect of $A+B$
  versus $P$ is
\begin{equation*} 
  \delta_{A+B,P} = \delta_{A,P} + \delta_{B,P} + \lambda_{AB}
\end{equation*}
with $\lambda_{AB} = \log(\mbox{\textit{IR}}_{AB})$. Following
Thorlund and Mills \cite{Thorl:Mills:2012}, we set
$\mbox{\textit{IR}}_{AB}=1.5$.

\item[(c)] \textit{The additivity assumption for one intervention is
  strongly violated with a relevant synergistic interaction}. \newline
  The same model equation is used as for mild violation of additivity,
  however, the interaction ratio is larger in this scenario,
  $\mbox{\textit{IR}}_{AB}=2.0$.
\end{itemize}

Scenario A assumes additivity for all combined interventions in the
network, i.e., $A+B, A+C$, and $C+D$. Two variants of scenarios B and
C were considered. First, we assumed mild or strong violation of
additive effects for the combined intervention $A+B$ which are
labelled scenarios B1 and C1. Second, we assumed mild or strong
violation of additive effects for the combined intervention $C+D$
which are labelled scenario B2 and C2. Each scenario was repeated
$M=1000$ times. 

\FloatBarrier
\begin{table}[b]
\begin{center}
\caption{Overview of simulated scenarios \label{tab:simuset}}
\begin{tabular}{lrcc}
\hline
\textbf{Network geometry} & Well-connected network \\
Studies/Pairwise comparisons & $k=28$ \\
\hline 
\textbf{Interventions} & $n=8$\\
                       & single: $A,B,C,D,$ \\
                       & combined: $A+B,A+C,C+D,$ \\
                       & reference: placebo $P$ \\
\hline
\textbf{Additivity assumption on relative intervention effects} \\

Scenario A: Additive effects &
$\delta_{A+B,P} = \delta_{A,P} + \delta_{B,P}$ \\
&
$\delta_{A+C,P} = \delta_{A,P} + \delta_{C,P}$ \\
&
$\delta_{C+D,P} = \delta_{C,P} + \delta_{D,P}$ \\                                                
Scenario B: Mild violation of additivity assumption \\
B1: combined intervention $A+B$ &
$\delta_{A+B,P} = \delta_{A,P} + \delta_{B,P} + \lambda_{AB},
e^{\lambda_{AB}} = 1.5$ \\
B2: combined intervention $C+D$ &
$\delta_{C+D,P} = \delta_{C,P} + \delta_{D,P} + \lambda_{CD},
e^{\lambda_{CD}} = 1.5$ \\
Scenario C: Strong violation of additivity assumption \\
C1: combined intervention $A+B$ &
$\delta_{A+B,P} = \delta_{A,P} + \delta_{B,P} + \lambda_{AB}$,
$e^{\lambda_{AB}} = 2.0$ \\
C2: combined intervention $C+D$ &
$\delta_{C+D,P} = \delta_{C,P} + \delta_{D,P} + \lambda_{CD}$,
$e^{\lambda_{CD}} = 2.0$ \\
\hline 
\textbf{Heterogeneity}\\
No heterogeneity       & $\tau^2=0.00$ \\
Low heterogeneity      & $\tau^2=0.01$ \\
Moderate heterogeneity & $\tau^2=0.10$ \\
\hline
\textbf{Inconsistency} & No inconsistency \\
\hline  
\textbf{Other simulation parameters} \\  
True relative intervention effects &
$e^{\delta_{A,P}}= 1.40$, $e^{\delta_{B,P}} = 1.20$, $e^{\delta_{C,P}}= 2.30$, and
$e^{\delta_{D,P}}= 1.50$ \\
Baseline probability & $p_{P}=0.1$ \\
Patients per study arm & $n_i\sim \mathcal{U}(50, 200)$ \\
Iterations & $M=1000$ \\
\hline
\end{tabular}
\end{center}
\end{table}
\FloatBarrier

\subsection{Create disconnected networks \label{sec:sim:gen:disc}}

In our simulation study, we started by simulating connected networks
and afterwards, we artificially constructed all possible disconnected
networks as described in Section \ref{sec:disc:generate}. The
well-connected network in our simulation study does not have any
intervention only connected to the reference $P$. Accordingly, the
smallest main subnetwork in the simulation study is any single active
intervention vs $P$. Additional interventions can be added to this
small main subnetwork. In each simulation run, we randomly selected
one disconnected network from the set of possible disconnected
networks.

\subsection{Generation of simulated data}

The generation of binary data was similar to Kiefer
et. al.\ \cite{kiefer2020} For each study $i = 1, \dots,
k$, we generated study-specific log-odds ratios $d_{iX,Y}$ from a
normal distribution with mean $d_{X,Y}$ and between-study variance
$\tau^2$ with $X,Y \in \{A,B,C,D,A+B,A+C,C+D,P\}$, $X \ne Y$,
representing the set of possible interventions. The baseline
probability for placebo was set to $0.1$. Study-specific
probabilities for intervention $t\in \{A,B,C,D,A+B,A+C,C+D\}$ were
calculated by $$p_{it}=
\frac{0.1\exp(d_{it,P})}{1-0.1(1-\exp(d_{it,P}))}$$ with
log-odds ratios $d_{it,P}$.

For each study, we generated equal arm sample sizes $n_i$ from
a discrete uniform distribution assuming values from $50$ to
$200$. For study arms 1 and 2, we generated the number of events
$e_{it_1}$ and $e_{it_2}$, $t_1, t_2 \in \{A,B,C,D,A+B,A+C,C+D, P\},
t_1 \ne t_2$, randomly from a binomial distribution with parameters
$n_i$ and $p_{it_1}$ or $p_{it_2}$. If the simulated number of events
$e_{it_1}$ or $e_{it_2}$ was zero, we added the value $0.5$ to both
event numbers.

\subsection{Simulation performance}

We first calculated the mean squared error (MSE) and the coverage probability (CP) of the relative intervention effect, for each pair of the seven active interventions and placebo.
Then, we calculated the average MSE and CP to summarize the properties of (C)NMA model fit. 
For each setting (connected or disconnected network, scenarios A to
C2) and value of $\tau^2$, we calculated the average MSE as

\begin{equation*} 
  \MSE = \frac{1}{M} \sum_{m=1}^M
  \frac{1}{7} \left \| \hat\bfdelta_{m} - \bfdelta \right \|^2
\end{equation*}

where $\hat\bfdelta_m = (\hat\delta_{m, A,P}, \hat\delta_{m, B,P},
\hat\delta_{m, C,P}, \hat\delta_{m, D,P}, \hat\delta_{m, A+B,P},
\hat\delta_{m, A+C,P}, \hat\delta_{m, C+D,P})$ denotes the vector with
estimated relative intervention effects in iteration $m, m = 1, \dots,
M$, and $\hat\bfdelta$ denotes the corresponding true effects. The
division by 7 refers to the seven baseline parameters in this network.

The average coverage probability was similarly defined

\begin{equation*} 
  \CP = \frac{1}{M} \sum_{m=1}^M \frac{1}{7} \ 
\left \| \bf{1}_{\hat\bfdelta_{m, L} \leq \bfdelta \leq \hat\bfdelta_{m, U}} \right \|
\end{equation*}

where $\hat\bfdelta_{m, L}$ and
$\hat\bfdelta_{m, U}$ are vectors with lower
and upper 95\% confidence limits calculated in iteration $m, m = 1,
\dots, M$ and $\bf{1}$ is a vector of indicator functions.

\subsection{Software implementation}

The simulation study was performed using the statistical software R
\cite{RDev:lang}. (C)NMA models were fitted with R package
\textbf{netmeta} using function \texttt{netcomb} for connected
networks and \texttt{discomb} for disconnected
networks. \cite{Ruck:Krah:Koni:Efth:netmeta}

\section{Results}\label{sec:results}

\subsection{CNMA model selection in Cochrane data set}\label{sec:results:nausea}

\subsubsection{Connected network}

A standard NMA of the Cochrane data set does not show any substantial
between-study heterogeneity / inconsistency ($Q = 44.80, \df = 46, p =
0.5227$). In the additive CNMA model which estimates 17 component
effects (including placebo) the between-study heterogeneity /
inconsistency is much larger ($Q = 103.53, \df = 55, p < 0.0001$). The
difference in $Q$ statistics between additive CNMA and standard NMA is
highly statistical significant $(Q_{\mbox{\textit{diff}}} = 58.74, \df
= 9, p < 0.0001)$, showing that the additivity assumption is not
justified for all observed treatment combinations. Accordingly, we
used the forward CNMA model selection procedure to add interaction
terms and to relax the additivity assumption.

The network has 11 combinations of two interventions, however, one
component (vest) was only evaluated in the combination onda +
vest. Accordingly, the interaction onda$*$vest cannot be distinguished
from the net effect of vest, leaving ten potential 2-way interaction
terms. Starting from the additive CNMA model, we added a single 2-way
interaction term to the model (Table A2). All ten interaction models
led to a reduction of $Q$ with the largest reduction for the
interaction onda$*$scop ($Q = 53.70, \df = 54, p = 0.4860$). This
interaction CNMA model is preferred over the additive CNMA model
according to the AIC criterion ($p < 0.0001$).

In the next selection step, all 45 combinations of two 2-way
interactions were considered. Only nine models with two 2-way
interactions led to a further reduction of $Q$ (Table A2). All nine
models included the interaction onda$*$scop selected in the first step
plus one additional interaction term. The largest reduction of $Q$ was
observed for the interactions onda$*$scop + apre$*$scop $(Q = 50.19,
\df = 53, p = 0.5841)$. This model was preferable to the CNMA model
with one 2-way interaction according to the AIC criterion $(p =
0.0611)$.

In total, ten of 120 combinations of three 2-way interactions further
decreased $Q$. All ten models included the interaction onda$*$scop,
however, only 8 of 10 models included the interaction apre$*$scop
which was selected in the second step. Three of the 10 models included the
interaction meto$*$trop which was also part of the model with the
smallest heterogeneity/inconsistency: onda$*$scop + apre$*$scop +
meto$*$trop $(Q = 47.71, \df = 52, p = 0.6432)$. Based on the AIC
criterion, this combination of three 2-way CNMA interactions was
preferable to a model with two 2-way interactions ($p =
0.1197$). Overall, the results of this selected interaction CNMA model are
very similar to the standard NMA (Figure A1). Only for the combination
meto + scop versus plac, the standard NMA reports a non-significant
result (RR $= 1.65$ $[0.66; 4.12]$) while the selected interaction
CNMA model estimates a stronger, significant effect (RR $= 2.28$ $[1.74;
  2.99]$). In contrast, results for the additive CNMA are different
for several comparisons, especially, onda + scop vs plac or scop vs
plac (Figure A1).

\subsubsection{Disconnected networks}

We used the strategy described in Section \ref{sec:disc:generate} to
identify additional disconnected networks including all 27
interventions. Eight additional disconnected networks could be
identified using R function \texttt{disconnect\_additional} (see
GitLab repository). These networks, which differ substantially in the
number of the included studies and the number of pairwise comparisons, were sorted by
decreasing number of pairwise comparisons, the decreasing number of
studies and decreasing number of pairwise comparisons in the main
subnetwork (Table \ref{tab:res:nausea}; Figure A2).

The largest disconnected network contains 55 of 57 studies and 87 of
89 pairwise comparisons. The main subnetwork was very similar to the
connected network only excluding the two studies comparing onda with
apre (discarded to generate a disconnected network) and the single
study comparing apre with apre + scop (auxiliary subnetwork). On the
other hand, disconnected network 9 only included 36 studies with 54
comparison distributed over three subnetworks.

\begin{table}[tb]
  \caption{Results of CNMA model selection for disconnected networks
    in Cochrane data set with number of studies $k$ and number of
    pairwise comparisons $m$ \label{tab:res:nausea}}
  \begin{tabular}{lccccccc}
    \hline
    & & &   One 2-way & Combination of two & Combination of three \\
    \rb{Network} & \rb{$k$} & \rb{$m$} & interaction & 2-way interactions & 2-way interactions \\ \hline
    Connected      & 57 & 89 & onda$*$scop & onda$*$scop + apre$*$scop\phantom{$^\star$} &
    onda$*$scop + apre$*$scop + meto$*$trop\phantom{$^\spadesuit$} \\
    Disconnected 1 & 55 & 87 & onda$*$scop & onda$*$scop + apre$*$scop\phantom{$^\star$} &
    onda$*$scop + apre$*$scop + meto$*$trop\phantom{$^\spadesuit$} \\
    Disconnected 2 & 41 & 61 & onda$*$scop & onda$*$scop + dexa$*$trop$^\star$ &
    onda$*$scop + dexa$*$gran + dexa$*$trop$^\spadesuit$ \\
    Disconnected 3 & 39 & 59 & onda$*$scop & onda$*$scop + dexa$*$trop$^\star$ &
    onda$*$scop + dexa$*$gran + dexa$*$trop$^\spadesuit$ \\
    Disconnected 4 & 39 & 59 & onda$*$scop & onda$*$scop + dexa$*$trop$^\star$ &
    onda$*$scop + dexa$*$gran + dexa$*$trop$^\spadesuit$ \\
    Disconnected 5 & 40 & 58 & onda$*$scop & onda$*$scop + dexa$*$trop$^\star$ &
    onda$*$scop + dexa$*$gran + dexa$*$trop$^\spadesuit$ \\
    Disconnected 6 & 37 & 57 & onda$*$scop & onda$*$scop + dexa$*$trop$^\star$ &
    onda$*$scop + dexa$*$gran + dexa$*$trop$^\spadesuit$ \\
    Disconnected 3 & 38 & 56 & onda$*$scop & onda$*$scop + dexa$*$trop$^\star$ &
    onda$*$scop + dexa$*$gran + dexa$*$trop$^\spadesuit$ \\
    Disconnected 8 & 38 & 56 & onda$*$scop & onda$*$scop + dexa$*$trop$^\star$ &
    onda$*$scop + dexa$*$gran + dexa$*$trop$^\spadesuit$ \\
    Disconnected 9 & 36 & 54 & onda$*$scop & onda$*$scop + dexa$*$trop$^\star$ &
    onda$*$scop + dexa$*$gran + dexa$*$trop$^\spadesuit$ \\ \hline
  \end{tabular}
  
  $^\star$ Second best model: onda$*$scop + apre$*$scop
  \\
  $^\spadesuit$ Second best model: onda$*$scop + apre$*$scop + meto$*$trop
\end{table}

Table \ref{tab:res:nausea} provides the results of the model selection
process which was applied to all disconnected networks.  After fitting
additive CNMA models, we started the model selection by adding a
single 2-way interaction. In all disconnected networks, the
interaction onda$*$scop minimized $Q$ and was selected according to
the AIC criterion. All observed combinations of two and three 2-way
interaction CNMA models were considered. According to the AIC
criterion, a combination of three 2-way interaction CNMA models was
chosen for all disconnected networks. The combination onda$*$scop +
apre$*$scop + meto$*$trop was selected for the connected network and
disconnected network 1. The combination onda$*$scop + dexa$*$gran +
dexa$*$trop was selected for disconnected networks 2-9. This
combination was also the second best model for the connected and
disconnected network 1 (Table A3).
  
We exemplify the results of the standard NMA and the selected
interaction CNMAs of the connected network or disconnected networks by
looking at the relative intervention effects amis, apre, apre + scop,
palo, ramo compared to placebo (Figure \ref{fig:forest-detail}).  The
full forest plot is provided in Supplementary Figure A3.  (C)NMA
results for the comparison amis versus plac are identical for all
models (Figure \ref{fig:forest-detail}). The intervention amis is only
directly compared to plac (Figure \ref{fig:networkplot}) and therefore
always part of the main subnetwork in disconnected networks with
placebo as reference. The relative intervention effect of apre + scop
or apre versus plac is very similar to the standard NMA and the first
selected CNMA of the connected network (Figure
\ref{fig:forest-detail}). The estimate of apre + scop versus plac is
misleading in the second selected CNMA of the connected network,
probably because the interaction apre$*$scop is not part of this
model. In disconnected networks, this effect is either inestimable
(disconnected networks 1, 3, 6, 7, 9) or misleading (disconnected
networks 2, 4, 5, 8). For the disconnected networks 1, 3, 6, 7,  and 9,
there is only one study comparing apre + scop versus apre, and it is
disconnected from the rest of the network (Figure A1). The component
effect apre cancels out due to the additivity assumption, and
therefore, the relative intervention effect for apre or apre + scop
versus plac is inestimable. For disconnected networks 2, 4, 5, and 8,
the relative intervention effects of apre or apre + scop versus plac
are not reliably estimated as intervention apre or apre + scop is not
part of the main subnetwork containing plac. Results for the
comparisons palo versus plac and ramo versus plac are very similar for
the standard NMA, selected CNMAs of the connected network and some
disconnected networks. Markedly wider confidence intervals are
observed for relative effects of palo versus plac (disconnected
networks 4, 6, 8, 9) and ramo versus plac (disconnected networks 5,
7-9). In these disconnected networks, the intervention palo or ramo is
not part of the main subnetwork containing plac.

In total, 6 out of 26 confidence intervals of relative intervention
effects for disconnected networks 2-9 do not include the estimate from
the standard NMA (e.g., apre vs plac, apre + scop vs plac) (Figure
A2).

\subsection{CNMA model selection for simulated connected networks}

For each scenario and value of $\tau^2$, 1000 connected networks were
simulated and the CNMA model selection was performed. Table
\ref{tab:sim:conn} summarizes simulation results for connected networks.

The number of simulations rejecting the additivity assumption at a
significance level of 5\% is given in the column
$n_{\mbox{\textit{diff}}}$. This test is based on
$Q_{\mbox{\textit{diff}}}$, i.e., the difference in $Q$ statistics
between additive CNMA and standard NMA. Under scenario A, the
percentage of significant results increases from 3.5\% to 18.8\% with
increasing $\tau^2$. Accordingly, the test for additivity is too
conservative for $\tau^2 = 0$ and too liberal for $\tau^2 = 0.1$. For
scenarios B and C, results in column $n_{\mbox{\textit{diff}}}$
describe the power to detect the violation of additivity. Overall, the
power is low for scenario B (25.7\% to 42.0\%) and moderate for
scenario C (65.6\% to 73.8\%).

\FloatBarrier
\begin{figure}[htb!]
  \centerline{\includegraphics[width=12cm,angle=0]{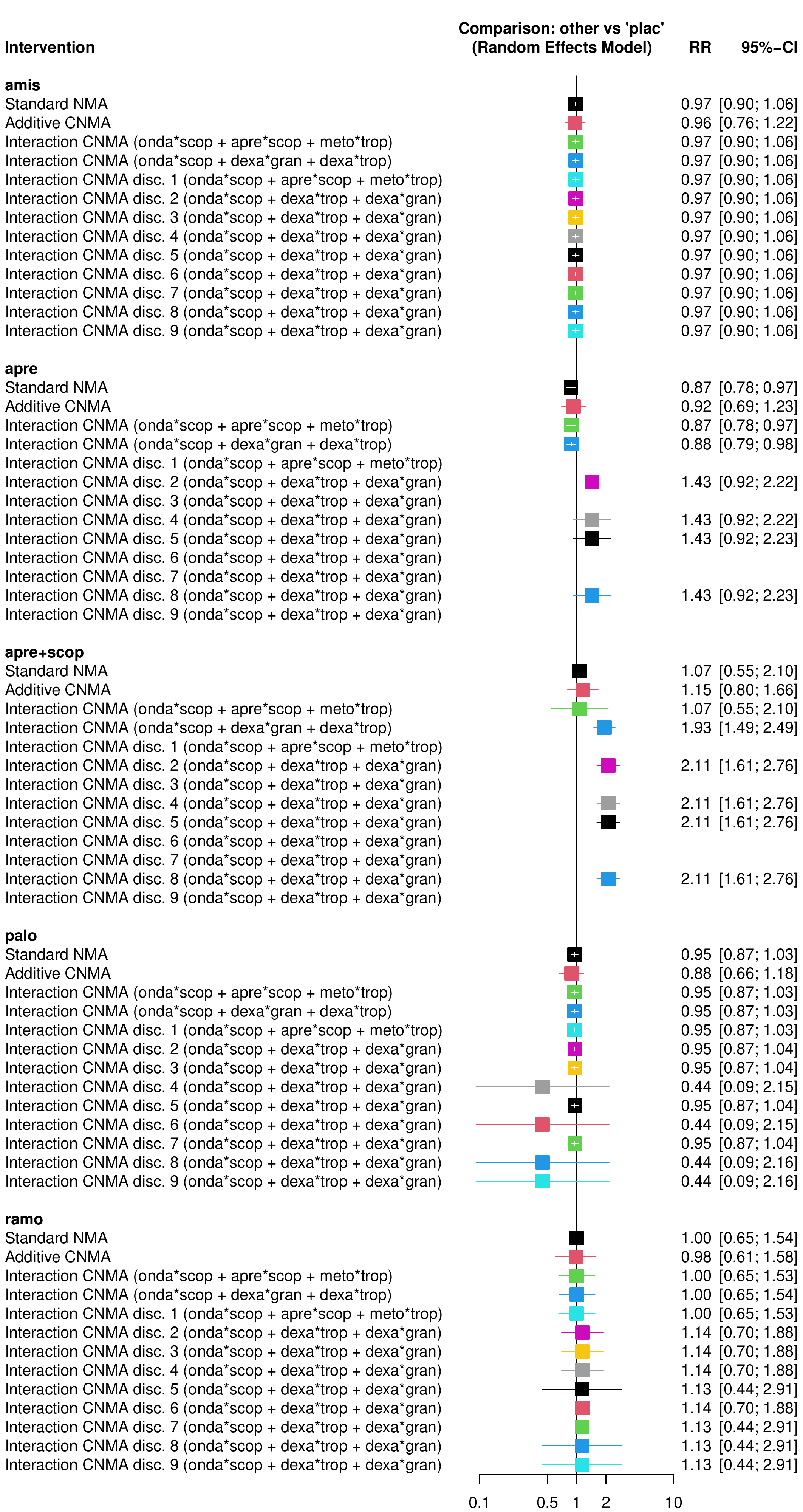}}
  \caption{Forest plot of the relative effects of interventions amis,
    apre, apre + scop, palo, ramo with placebo from standard NMA and
    the selected CNMAs for the connected and each disconnected
    network.\label{fig:forest-detail}}
\end{figure}
\FloatBarrier

CNMA model selection worked best for no heterogeneity (Table
\ref{tab:sim:conn}). In general, the number of simulations selecting
the correct model is decreasing with increasing heterogeneity.  If the
additivity assumption holds (scenario A), the correct additive CNMA
model was only selected in the majority of simulations in the CNMA
model selection for no or low heterogeneity (69.3\% and 65.2\%). For
moderate heterogeneity, less than 50\% of simulations selected the
additive CNMA model. For mild violation of the additivity assumption
(scenario B1 and B2), the correct interaction CNMA model was selected
in less than 50\% of simulations (34.6\% to 45.0\%). For strong
violation of additivity (scenarios C1 and C2), the correct interaction
CNMA model was always selected in the majority of simulations (54.6\%
to 74.6\%).

\newcommand{\myst}{{$^\star$}}
\newcommand{\pyst}{\phantom{\myst}}
\FloatBarrier
\begin{table}[h]
  \caption{Selected CNMA models in simulations of connected
    networks \label{tab:sim:conn}}
  \begin{tabular}{lrccccccc} \hline
     & & & \multicolumn{3}{c}{CNMA with 2-way interaction} & \multicolumn{3}{c}{CNMA with two 2-way interactions} \\
    \rb{Scenario} & \rb{$n_{\mbox{\textit{diff}}}$} & \rb{Additive CNMA} &
   $A$$*$$B$ & $A$$*$$C$ & $C$$*$$D$ & $A$$*$$B$ + $C$$*$$D$ & $A$$*$$B$ + $A$$*$$C$ & $A$$*$$C$ + $C$$*$$D$ \\
    \hline
    \multicolumn{9}{l}{No heterogeneity ($\tau^2=0.00$)} \\
    A  &  35 & \textbf{693} &         119  &  94 &          84  &  3      &  2      &  5      \\
    B1 & 257 &         324  & \textbf{436} & 148 &          61  & 16\myst & 12\myst &  3      \\
    B2 & 298 &         325  &          45  & 154 & \textbf{436} & 17\myst &  8      & 15\myst \\
    C1 & 687 &          61  & \textbf{746} &  99 &          17  & 34\myst & 28\myst & 15\pyst \\
    C2 & 732 &          51  &          13  & 104 & \textbf{742} & 38\myst &  7      & 45\myst \\
    \hline
    \multicolumn{9}{l}{Low heterogeneity ($\tau^2=0.01$)} \\
    A  &  52 & \textbf{652} &         136  &  98 &         108  &  3      &  1      &  2      \\
    B1 & 281 &         337  & \textbf{426} & 136 &          59  & 16\myst & 18\myst &  8      \\
    B2 & 284 &         317  &          49  & 133 & \textbf{450} & 25\myst & 10      & 16\myst \\
    C1 & 676 &          75  & \textbf{704} & 109 &          24  & 29\myst & 47\myst & 12\pyst \\
    C2 & 738 &          70  &          13  &  98 & \textbf{716} & 45\myst &  7\pyst & 51\myst \\
    \hline
    \multicolumn{9}{l}{Moderate heterogeneity ($\tau^2=0.10$)} \\
    A  & 188 & \textbf{458} &         188  & 142 &         179  & 14\pyst & 13      &  6      \\
    B1 & 392 &         277  & \textbf{346} & 178 &         132  & 34\myst & 17\myst & 16\pyst \\
    B2 & 420 &         262  &         122  & 195 & \textbf{356} & 28\myst & 13\pyst & 24\myst \\
    C1 & 656 &         117  & \textbf{550} & 135 &          53  & 65\myst & 64\myst & 16\pyst \\
    C2 & 702 &          93  &          48  & 161 & \textbf{546} & 65\myst & 18\pyst & 69\myst \\
    \hline
  \end{tabular}
  
  The correctly chosen model is printed in \textbf{bold}.
  
  \myst Combination of two 2-way interactions includes the correct
  interaction.
  
  $n_{\mbox{\textit{diff}}}$: number of networks with significant
  $p$-value for $Q_{\mbox{\textit{diff}}}$, where
  $Q_{\mbox{\textit{diff}}}$ is the $Q$ statistic for the difference
  between additive CNMA and standard NMA model.
\end{table}
\FloatBarrier

\begin{figure}[h]
  \centerline{\includegraphics[width=12cm,angle=0]{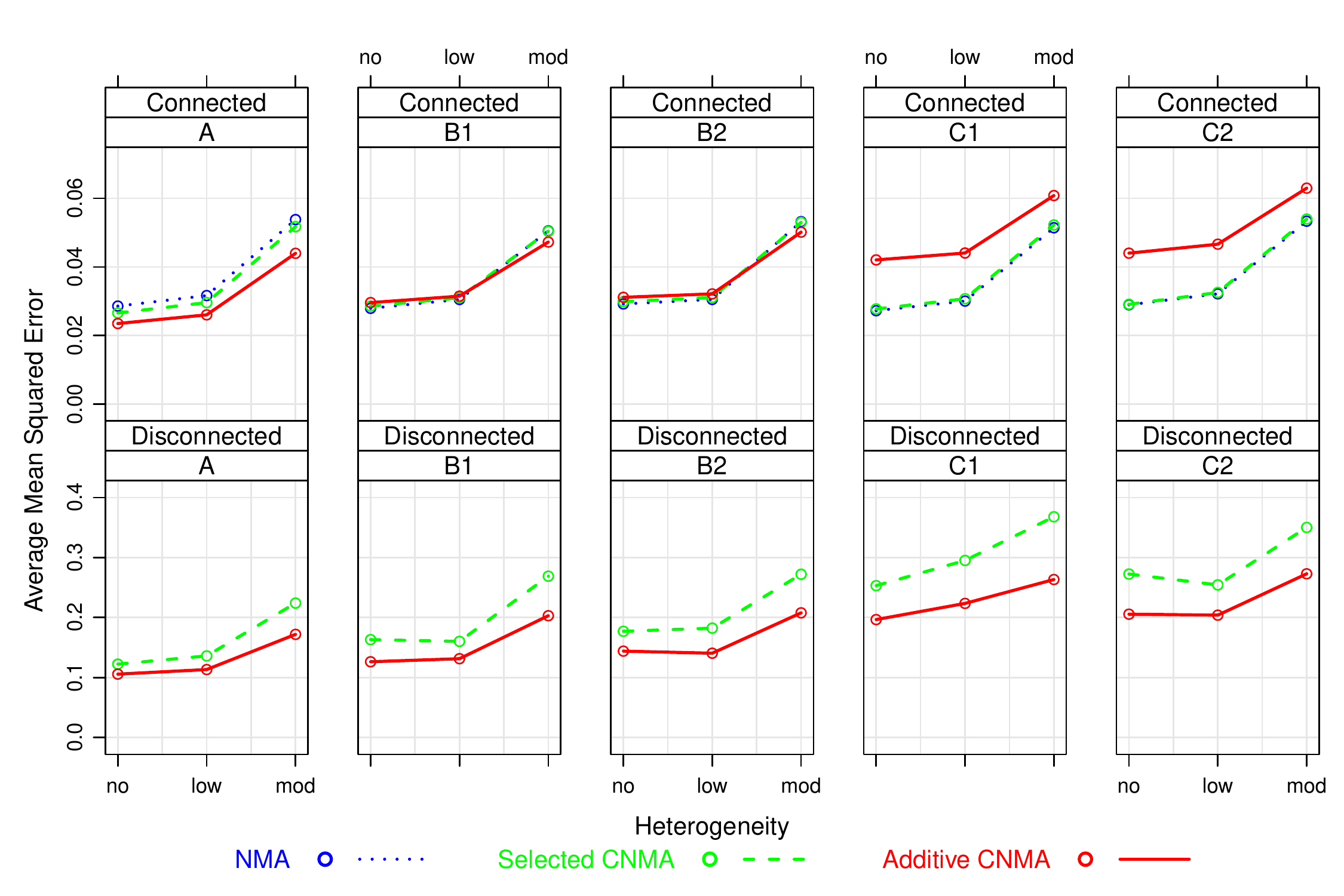}}
  \caption{Average mean squared errors for simulated connected
    networks (top panel) and disconnected networks (bottom
    panel). Different scales are used on the y-axis due to the large
    differences in MSEs for connected and disconnected
    networks. \label{fig:sim:mse}}
\end{figure}

Figure \ref{fig:sim:mse}, top panel, provides the average mean squared
errors for the simulated connected networks. Under scenario A, average
MSEs of the additive CNMA model is slightly smaller than the MSEs for
selected iCNMA and standard NMA. All models perform on average equally
for mild violation of the additivity assumption (scenarios B1 and B2)
while selected iCNMA and standard NMA models perform better for a large
violation of the additivity assumption (scenarios C1 and C2). Overall,
the selected iCNMA and standard NMA have very similar average
MSEs. Furthermore, average MSEs get larger with increasing
heterogeneity. Figure A4 shows that MSEs of the relative intervention effects
are comparable for selected iCNMA and standard NMA model,
while the additive CNMA model has inferior results for several
intervention estimates for a strong violation of the additivity
assumption: $B$ vs $P$, $C$ vs $P$ and $A+B$ (scenario C1) and $A$ vs
$P$ and $D$ vs $P$ (scenario C2).

\begin{figure}[h]
  \centerline{\includegraphics[width=12cm,angle=0]{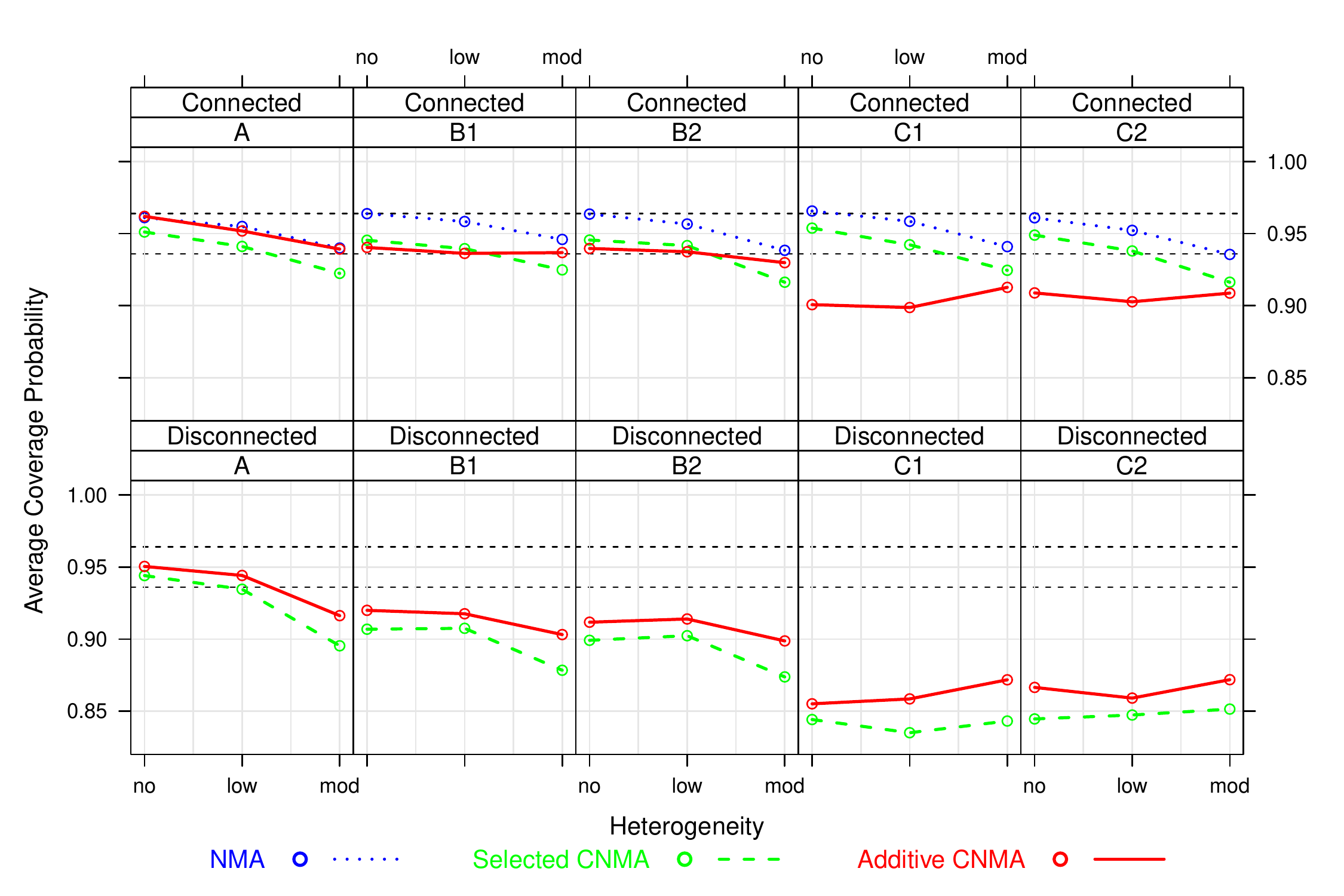}}
  \caption{Average coverage probabilities for simulated connected
    networks (top panel) and disconnected networks (bottom
    panel). Dashed lines denote the 95\% Monte-Carlo limits
    $(L,U)=[0.936; 0.964]$ for CP. \label{fig:sim:cp}}
\end{figure}

Figure \ref{fig:sim:cp}, top panel, provides the average coverage
probabilities for the simulated connected networks. Under scenario A,
the average coverage probability is decreasing with increasing
heterogeneity for all three models. Average coverage probabilities are
very similar for standard NMA and additive CNMA. Coverage
probabilities lie within the 95\% Monte-Carlo limits, with exception
of the selected interaction CNMA model for moderate heterogeneity. For
mild and strong violation of the additivity assumption, only the
coverage probability of the standard NMA model always falls within the
95\% Monte-Carlo limits for any value of $\tau^2$. Coverage
probabilities for the additive model are well below the lower
Monte-Carlo limit for scenarios C1 and C2. Figure A5 shows that
coverage probabilities of the relative intervention effects are in general somewhat smaller for the selected iCNMA model
compared to the standard NMA model, however, not dramatically
different. For the additive CNMA model, coverage probabilities are
much smaller than the Monte-Carlo limits for the intervention
estimates with large MSE values.

\FloatBarrier
\FloatBarrier

\FloatBarrier
\FloatBarrier

\subsection{CNMA model selection for simulated disconnected networks}

Table \ref{tab:sim:disc} summarizes simulation results for
disconnected networks. In comparison to connected networks, no test of
the additivity assumption is available, as a standard NMA model cannot
be estimated in a disconnected network.

CNMA model selection did not work in our simulation in the simulated
disconnected networks. Despite using the rather liberal Akaike
criterion, the additive CNMA model is selected in the majority of
simulations under mild violation of additivity (58.3\% to 70.4\%) and
in a large proportion of simulations under strong violation of the
additivity assumption (46.2\% to 51.5\%). The correct interaction CNMA
model is only selected in 12.0\% to 15.2\% under scenarios B1 and B2
and 21.6\% to 26.3\% under scenarios C1 and C2.

Figure \ref{fig:sim:mse}, bottom panel, provides the average mean
squared errors for the simulated disconnected networks. For all
scenarios, MSEs for disconnected networks are much larger than for
connected networks. Furthermore, results for the selected iCNMA are
always worse than for the additive CNMA, even under scenarios C1 and C2
with strong violation of the additivity assumption. Average coverage
probabilities for the simulated disconnected networks are given in
Figure \ref{fig:sim:cp}, bottom panel. Coverage probabilities only
fall into the 95\% Monte-Carlo limits for scenario A and no or low
heterogeneity. Again, average coverage probabilities are always worse
for the selected iCNMA. This general pattern is also observed for MSEs
and CPs of the relative intervention effects (Figures A6 and A7).

\FloatBarrier
\begin{table}[h]
\begin{center}
 \caption{Selected CNMA models in simulations of disconnected
   networks \label{tab:sim:disc}}
 \begin{tabular}{lrccccccccc}
   \hline
   & & \multicolumn{3}{c}{CNMA with 2-way interaction} & \multicolumn{3}{c}{CNMA with two 2-way interactions} \\
   \rb{Scenario} & \rb{Additive CNMA} &
   $A$$*$$B$ & $A$$*$$C$ & $C$$*$$D$ & $A$$*$$B$ + $C$$*$$D$ & $A$$*$$B$ + $A$$*$$C$ & $A$$*$$C$ + $C$$*$$D$ \\
   \hline
   \multicolumn{8}{l}{No heterogeneity ($\tau^2=0.00$)} \\
   A  & \textbf{834} &          52  &  55 &          57  & 1\pyst & 0\pyst & 1\pyst \\
   B1 &         698  & \textbf{121} & 120 &          57  & 1\myst & 1\myst & 2\pyst \\
   B2 &         693  &          79  &  93 & \textbf{131} & 1\myst & 3\pyst & 0\myst \\
   C1 &         511  & \textbf{227} & 158 &          98  & 2\myst & 2\myst & 2\pyst \\
   C2 &         469  &         107  & 179 & \textbf{237} & 2\myst & 2\pyst & 4\myst \\
   \hline
   \multicolumn{8}{l}{Low heterogeneity ($\tau^2=0.01$)} \\
   A  & \textbf{809} &          51  &  72 &          68  & 0\pyst & 0\pyst & 0\pyst \\
   B1 &         704  & \textbf{120} &  94 &          79  & 1\myst & 1\myst & 1\pyst \\
   B2 &         677  &          83  &  99 & \textbf{138} & 1\myst & 1\pyst & 1\myst \\
   C1 &         515  & \textbf{216} & 151 &         111  & 6\myst & 0\myst & 1\pyst \\
   C2 &         480  &         101  & 150 & \textbf{263} & 1\myst & 0\pyst & 5\myst \\
   \hline
   \multicolumn{8}{l}{Moderate heterogeneity ($\tau^2=0.10$)} \\
   A  & \textbf{664} &         116  & 110 &         103  & 3\pyst & 2\pyst & 2\pyst \\
   B1 &         598  & \textbf{148} & 131 &         115  & 5\myst & 1\myst & 2\pyst \\
   B2 &         583  &         137  & 119 & \textbf{152} & 2\myst & 4\pyst & 3\myst \\
   C1 &         462  & \textbf{220} & 171 &         137  & 6\myst & 3\myst & 1\pyst \\
   C2 &         485  &         130  & 154 & \textbf{216} & 8\myst & 4\pyst & 3\myst \\
   \hline
 \end{tabular}
 \\ The correctly chosen model is printed in bold. \myst The chosen
 combination of two 2-way interactions includes the correct
 interaction. \\
\end{center}
\end{table}
\FloatBarrier

\section{Discussion}\label{con}

In this article, we introduce a model selection strategy for component
network meta-analysis that can be used in connected or disconnected
networks. In addition, we describe a procedure to create disconnected
networks in order to evaluate the properties of the model selection for
both connected and disconnected networks. We apply the methods to
investigate their performance to Cochrane review
data and simulated data.

In connected networks, it is always possible to contrast the results of
standard NMA with additive or interaction CNMA models. Accordingly,
the application of CNMA models should always be accompanied by a
statistical test to assess additivity. We used the difference in $Q$
statistics between additive or interaction CNMA and standard NMA
model.

The application of the additivity test to the Cochrane data set
suggests that the additivity assumption does not hold. The network has
eleven combinations of two interventions, however, only ten 2-way
interactions are estimable. Among the ten 2-way interactions, the
interaction onda*scop was selected first. The selected CNMA model
includes three 2-way interactions (onda*scop + apre*scop + meto*trop)
and its results roughly agree with those of the standard NMA. For the
disconnected networks in the Cochrane data set, we observed that the
estimates can be sometimes similar, sometimes very different, or even
inestimable, depending on the network structure. A clinical
interpretation of the identified interactions was not the main focus of
this work. We think that this would be futile for the
general outcome of any adverse events. More specific outcomes should
be considered to give a clinical meaning to interactions in
CNMA models.

The results for the performance of CNMA models are in agreement with
the previous simulation study conducted for connected networks by
Thorlund and Mills \cite{Thorl:Mills:2012}. According to our
simulation study, the test for additivity with Q statistic of the
difference between the CNMA and NMA model only has sufficient power
for a strong violation of additivity. We found that the Q statistic
tends to stick to additivity, even if it is strongly violated. As
expected, the performance of the model selection procedure for the
connected networks deteriorates with increasing heterogeneity. We
conclude that the additive CNMA can be an alternative to NMA if the
additivity assumption holds and, in general, the selected CNMA and
standard NMA provide similar estimates.

The Q statistic of the difference between the CNMA and NMA model is
not available for disconnected networks, as standard NMA cannot be
fitted.  Simulation results are worse for the selected CNMA compared
to the additive CNMA for disconnected networks.  Our simulation
results show that re-connecting disconnected networks with CNMA models
is possible only when additivity can be safely assumed. For
disconnected networks, we recommend using additive CNMA only if strong
clinical arguments for additivity exist. Otherwise, the subnetworks
should be analyzed separately.

The test of additivity was found to have low power in some connected
networks. For disconnected networks, the additivity assumption cannot
be tested. New statistical techniques for the evaluation of additivity
assumption are required for both connected and disconnected networks.

Our simulation study shows that CNMA model selection works for
connected networks, but not for disconnected networks. Accordingly, we
see CNMA model selection as a useful tool for a connected network 
to evaluate potential interactions between the components of
multicomponent interventions. We considered CNMA model selection in
two directions, forward and backward.  As the aim of this simulation
study was to evaluate the performance of both connected and
disconnected networks, we decided to use only forward selection, to
achieve a satisfactory model fit whilst keeping much of the
connectivity that is given by the additive
model. \cite{Rucke:Schmi:Schwa:2020}. Forward selection tended to
select sparse (often additive) CNMA models for disconnected networks
even if additivity was mildly or strongly violated. We only considered
2-way interactions in our simulation, however, the selection procedure
could also be used with 3-way or higher interactions.

One limitation of our simulation study is that conclusions depend on
the scenarios considered. We simulated a network of interventions with
eight interventions and 28 two-arm studies, assuming consistency. We
also implemented the forward CNMA model selection process in
simulations with the Akaike Information criterion. It is unclear
whether different network structures, model assumptions, or a
different design or strategy for the model selection process would
lead to different conclusions. There is no guarantee that CNMA models
behave similarly under different simulation designs.

Although the use of NMA has considerably increased over the last
decade, CNMA has not been widely used, but there is an increased
clinical interest in the evaluation of multicomponent interventions
and we expect an increase in its use. CNMA models are now provided in
a frequentist framework, implemented in the R package
\textbf{netmeta}. This simulation study provides guidance for CNMA
model selection, pointing at some challenges that should be addressed.

\bibliography{MainDoc}

\end{document}